\documentclass[pra,twocolumn,superscriptaddress,amsmath,amssymb,aps,10pt]{revtex4-1}

\usepackage{graphicx}
\usepackage{dcolumn}
\usepackage{bm}
\usepackage{braket}
\usepackage{hyperref}
\hypersetup{
    colorlinks=true,        
    linkcolor=blue,          
    citecolor=blue,        
    filecolor=magenta,      
    urlcolor=blue           
}

\usepackage{color} 				
\definecolor{myblue}{rgb}{0.4, 0.3, 0.7}
\definecolor{dmag}{rgb}{0.6,0.0,0.6}
\definecolor{dred}{rgb}{0.8,0,0}
\definecolor{gray}{rgb}{0.5,0.5,0.5}
\usepackage[normalem]{ulem} 	
\newcommand{\change}[1]{{{#1}}}	
\definecolor{pink}{rgb}{1,0,0.9}

\begin{document}

\title{ Anisotropy-driven magnetic phase transitions in SU(4)-symmetric Fermi gas\\ in three-dimensional optical lattices
}

\author{Vladyslav Unukovych}
\affiliation{Karazin Kharkiv National University, Svobody Sq. 4, 61022 Kharkiv, Ukraine}

\author{Andrii Sotnikov}
\affiliation{Karazin Kharkiv National University, Svobody Sq. 4, 61022 Kharkiv, Ukraine}
\affiliation{Akhiezer Institute for Theoretical Physics, NSC KIPT, Akademichna Str.~1, 61108 Kharkiv, Ukraine}

\date{\today}

\begin{abstract}
We study SU(4)-symmetric ultracold fermionic mixture in the cubic optical lattice with the variable tunneling amplitude along one particular crystallographic axis in the crossover region from the two- to three-dimensional spatial geometry. To theoretically analyze emerging magnetic phases and physical observables, we describe the system in the framework of the Fermi-Hubbard model and apply dynamical mean-field theory. We show that in two limiting cases of anisotropy there are two phases with different antiferromagnetic orderings in the zero temperature limit and determine a region of their coexistence. We also study the stability regions of different magnetically-ordered states and density profiles of the gas in the harmonic optical trap.   
\end{abstract} 

\maketitle

\section{Introduction}

In the rich domain of quantum many-body systems, one often deals with symmetry considerations. In particular, for lattice systems with discrete translational symmetries it is necessary to distinguish spatial symmetry of the physical system itself from the symmetry of the Hamiltonian governing its many-body properties. For spinful particles the intricate interplay between both ingredients can enhance or suppress quantum correlations, giving rise to magnetic long-range ordering, spin-liquid behavior, valence-bond states, as well as more exotic phenomena \change{\cite{Savary2017, Wu2010, Gorshkov2010NP, Caz2014RPP}}.

In this respect, many-body systems with high spin symmetries have become a very suitable test bed for in-depth studies of competing symmetry-related mechanisms.
The experiments with \change{ultracold} $^{87}$Sr and $^{173}$Yb atoms loaded into optical lattice, in particular, allow for realizations of SU($N$)-symmetric Hamiltonians up to $N=6$ and $N=10$, respectively (see, e.g., Refs.~\cite{Tai2010PRL,Ste2011PRA}). From the theory side, it is shown that for the fixed lattice geometry (e.g., on the square lattice) in the Heisenberg model one should observe suppression of the conventional magnetic order with the increase of $N$ \cite{Wu2010, Gorshkov2010NP, Caz2014RPP}.
\change{In this respect, the case of $N=4$ is especially interesting, since it is viewed as the one preceding the transition to the valence bond state with the further increase of $N$, i.e., still hosting local magnetic moments in the ground state.}

The similar effects are expected for the related Hubbard model, in particular, for the characteristic critical temperatures in units of the hopping amplitude~\cite{Golubeva2017PRB, Ibarra2023}, while in certain regimes, the magnetic correlations can be enhanced in high-symmetry cold atomic gases, if the entropy is used as a characteristic thermodynamic quantity \cite{Ozawa2018,Taie2022}.
\change{From this perspective, the systems with SU(4) spin symmetry have certain advantages from the experimental point of view \cite{Ozawa2018} and also due to the recently suggested additional cooling schemes \cite{Yamamoto2024}.
}

In case one keeps the spin symmetry of the model fixed to SU(2), e.g., by taking the spin-1/2 Heisenberg model, but changes the lattice geometry instead, the suppression of antiferromagnetic ordering can be observed already by changing the lattice geometry, for example, to triangular or kagome \cite{Leung1993,Huse2007}.
\change{This becomes especially relevant in view of strong experimental advances for ultracold Fermi gases with SU(2) symmetry in optical lattices reported recently (see, e.g., Refs.~\cite{Shao2024, Lebrat2024}).}
The aspects of geometric frustration remain crucial (also, becoming more intricate) for the SU(3)-symmetric spin models on these lattices as well \cite{Toth2010, Corboz2012}.

Within the current study, our specific goal is to explore how the continuous change between two-dimensional (2D) square and three-dimensional (3D) cubic lattice geometry affects magnetic ordering in the SU(4)-symmetric Hubbard model. 
\change{
The only former 2D case was studied in detail in Ref.~\cite{unukovych20214} in the framework of the dynamical mean-field theory \cite{Georges1996RMP}, which is one of the limiting cases of a broader class, i.e., the 2D-3D crossover region being in the focus of the current study. By using the developed approach, here we analyze how the change of lattice dimensionality}
affects critical behavior and experimentally-relevant local observables. In particular, we study the modulations in local filling of lattice sites by pseudospin components (i.e., antiferromagnetic correlations), regimes of potential coexistence of different magnetic phases, as well as real-space distributions of atomic densities in the presence of harmonic confinement.

\section{Model and Method}\label{sec.2}
We are interested in description of magnetic properties of four-component ultracold Fermi gas trapped in the cubic optical lattice, with the tunneling amplitudes equal each other in two spatial directions and the third one, which can be different from the other two (we denote the corresponding axis as $z$ below, i.e., $t_x=t_y\neq t_z$). On top of that, we choose that all four components are interacting equally, which can be experimentally realized with alkaline-earth(-like) atoms, in particular, \change{$^{87}$Sr \cite{Ste2011PRA} or $^{173}$Yb \cite{Ozawa2018}}. 

For sufficiently deep optical lattices, one can work within the framework of the Fermi-Hubbard model, with the local interactions between particles and tunneling restricted to the nearest-neighbor sites \cite{Gorshkov2010NP}:
\begin{align} \label{eq:Hubbard_Ham}
\hat{\cal H} = &-\sum \limits_{\langle i,j \rangle} \sum \limits_{\change{\sigma=1}}^{4}t_{ij} \hat{c}_{i \sigma}^{\dagger} \hat{c}_{j \sigma} + \sum \limits_{j} \sum \limits_{\sigma=1}^{4} \left(V_j-\mu\right) \hat{n}_{j \sigma}\nonumber \\
    & + U\sum \limits_{j} \sum \limits_{\sigma' > \sigma=1}^{4}  \hat{n}_{j \sigma} \hat{n}_{j \sigma'},
\end{align}
where $\hat{c}^{\dagger}_{i\sigma}$, $\hat{c}_{i\sigma}$, and $\hat{n}_{i\sigma}$ are the creation, annihilation, and number operators of particles in the pseudospin state~$\sigma$ on the site $i$, respectively. The notation $\langle i,j\rangle$ stands for all nearest-neighbor pairs, thus the first term describes tunneling from the site $j$ to $i$ with the amplitude $t_{ij}$, which can be either $t_x=t_y=t$ or $0\leq t_z \leq t$, depending on the direction. The second term determines filling of the lattice by atoms, where $V_j$ is the amplitude of the external potential acting on the lattice site $j$ and $\mu$ is the chemical potential. Unless it is specified separately, we consider a homogeneous system with $V_j=0$.
Meanwhile, the third term represents the onsite interaction between particles in different spin states. The Hamiltonian~\eqref{eq:Hubbard_Ham} is SU(4)-symmetric, i.e., remains invariant under action of the respective generators (generalized Gell-Mann matrices).

Below, we apply the numerical methodology directly to the Fermi-Hubbard model, but it is insightful to consider the limit of the strong interactions $U\gg t$. This limit allows us to more naturally introduce the ``magnetic'' terminology, as well as to better understand critical behavior at elevated temperatures in the system under study. By performing the Schrieffer-Wolff transformation of the Hamiltonian~\eqref{eq:Hubbard_Ham} and limiting the series by the second order in $t/U$, one arrives at the effective SU(4)-symmetric Heisenberg model,
\begin{align} \label{eq:Heis_Ham}
    \hat{\cal H}_{\rm eff}=\sum_{\langle i,j\rangle}\sum_{k=1}^{15}\change{J_{ij}}\hat{S}_{ki}\hat{S}_{kj}.
\end{align}
Here \change{$J_{ij}=4t_{ij}^2/U$} is the antiferromagnetic exchange coupling, \change{which can be either $4t^2/U$ or $4t_z^2/U$, depending
on the direction,} and the spin operator $\hat{S}_{ki}$ (acting at the site~$i$) is defined by the $k$-th generalized $4\times4$ Gell-Mann matrix~$\lambda_k$ ($k=1,...,15$) as $\hat{S}_{ki}=\frac{1}{2}\sum\limits_{\sigma,\sigma'=1}^4\hat{c}^{\dagger}_{i\sigma}\lambda^{}_{k\sigma\sigma'}\hat{c}^{}_{i\sigma'}$~\cite{Georgi1999}.

In order to analyze physical observables in different regimes of the introduced Fermi-Hubbard model~\eqref{eq:Hubbard_Ham}, we apply the dynamical mean-field theory (DMFT) \cite{Georges1996RMP} and its real-space extension (RDMFT) \cite{Hel2008PRL, Sno2008NJP, Sotnikov2015PRA, unukovych20214}. One of common approaches of these methods is to match the auxiliary Anderson impurity model to the original Fermi-Hubbard model, considering that each site (impurity) is surrounded by the multi-orbital ``external bath'',
\begin{align} \label{eq:Anderson_Ham}
    \hat{\cal H}_{A}=&-\sum_{l}\sum_{\sigma=1}^{4}\epsilon_{l\sigma}\hat{a}^{\dagger}_{l\sigma}\hat{a}_{l\sigma}+\sum_{l}\sum_{\sigma=1}^{4}V_{l\sigma}\left(\hat{a}^{\dagger}_{l\sigma}\hat{c}_{\sigma}+\hat{c}^{\dagger}_{\sigma}\hat{a}_{l\sigma}\right) \nonumber \\
    &-\mu\sum_{\sigma=1}^{4}\hat{n}_{\sigma}+U\sum_{\sigma'>\sigma=1}^{4}\hat{n}_{\sigma}\hat{n}_{\sigma'}.
\end{align}
Here $l$ corresponds to the bath orbital index, while $\hat{a}^{\dagger}_{l\sigma}$ and $\hat{c}^{\dagger}_{\sigma}$ are the creation operators of particle in the pseudospin state~$\sigma$ on the orbital $l$ and on the impurity, respectively. The same notation is assumed for the annihilation operators. The amplitudes $\epsilon_{l\sigma}$ and $V_{l}$ are the so-called Anderson parameters, which also determine the non-interacting Green's function. The latter is known as the Weiss Green's function $\mathcal{G}_{\sigma}(i\omega_n)$ and can be expressed as follows:
\begin{align} \label{eq:Weiss_func}
    \mathcal{G}_{\sigma}^{-1}(i\omega_n)=i\omega_n+\mu-\sum_{l}\sum_{\sigma=1}^{4}\frac{|V_{l\sigma}|^2}{i\omega_n-\epsilon_{l\sigma}},
\end{align}
where  $\omega_n=\pi(2n+1)k_BT$ is the Matsubara frequency for certain $n\in \mathbb{Z}$ and temperature $T$.

Another function necessary for the algorithm is the impurity Green's function $G_{imp,\sigma}(i\omega_n)$. In particular, by employing the exact diagonalization (ED) numerical technique to the Anderson impurity problem \cite{Caffarel1994PRL}, this function can be naturally expressed by using the Lehmann representation \cite{Atland-Simons}:
\begin{align} \label{eq:Lehmann_Green}
    \nonumber
   G_{imp,\sigma}(i\omega_n)=&\frac{1}{\mathcal{Z}}\sum_{s,s'} \frac{\langle s|\hat{c}_{\sigma}|s'\rangle \langle s'|\hat{c}^{\dagger}_{\sigma}|s\rangle}{i\omega_n + E_{s}-E_{s'}} \\
   &\times\left[e^{-E_{s}/k_BT}+e^{- E_{s'}/k_BT}\right],
\end{align}
with the partition function $\mathcal{Z}=\sum_{s}e^{-E_{s}/k_BT}$ and a complete set of the eigenstates $|s\rangle$ of the impurity model \eqref{eq:Anderson_Ham} with the corresponding eigenvalues $E_s$.

Both the impurity Green's function $G_{imp,\sigma}(i\omega_n)$ and the Weiss function $\mathcal{G}_{\sigma}(i\omega_n)$ can be related to the self-energy $\Sigma_{\sigma}(i\omega_n)$ via the Dyson equation,
 \begin{align} \label{eq:Dyson}
     \Sigma_{\sigma}(i\omega_n)= \mathcal{G}_{\sigma}^{-1}(i\omega_n)- G_{imp,\sigma}^{-1}(i\omega_n),
 \end{align}
which is necessary to define the interacting lattice Green's function $G_{\sigma}(i\omega_n)$. Naturally, the lattice Green's function and impurity Green's function should coincide within the DMFT approach. This sets the self-consistency condition and convergence criteria for the algorithm. 

In the case of describing the translationally-invariant many-body states, the lattice Green's function can be expressed as
\begin{align} \label{eq:lat_Green}
 G_{\sigma}(i\omega_n)=\int \frac{D(\varepsilon,t_z/t)}{i\omega+\mu-\varepsilon-\Sigma_{\sigma}(i\omega_n)}\;d\varepsilon,
\end{align}
while for the case of the two-sublattice cover (valid for bipartite lattice geometries and corresponding many-body states) \cite{Georges1996RMP}
\begin{align} \label{eq:lat_Green_twosub}
G_{\sigma}^{j}(i\omega_n)=\zeta_{\sigma}^{\bar{j}}\int \frac{D(\varepsilon,t_z/t)}{\zeta_{\sigma}^A\zeta_{\sigma}^B-\varepsilon^2}\;d\varepsilon.
\end{align}
Here index $j=A,B$ and its opposite $\bar{j}=B,A$ denotes the sublattice, $\zeta_{\sigma}^{j}=i\omega_n+\mu-\Sigma_{\sigma}^{j}$, and $D(\epsilon,t_z/t)$ is the density of states, which also depends on the spatial anisotropy of the lattice system. Similarly to Ref.~\cite{Golubeva2015PRA}, for each value of the anisotropy parameter $t_z/t$, we numerically evaluate elliptic integrals to obtain the corresponding noninteracting density of states $D(\varepsilon,t_z/t)$.

Let us explicitly describe the steps in the DMFT algorithm. 
By restricting to the ED impurity solver, one can approximate the Weiss function~\eqref{eq:Weiss_func} by summing over finite number of bath orbitals $l=\{2,...,n_s\}$ for each spin component~$\sigma$. Specifically, for the four-component system we limit ourselves to $n_s=4$ per each spin component, noticing that the results are almost identical to those obtained at $n_s=3$. Then, we calculate impurity Green's function $G_{imp,\sigma}(i\omega_n)$ with Eq.~\eqref{eq:Lehmann_Green} and apply it to find the self-energy via the Dyson equation~$\eqref{eq:Dyson}$. The self-energy $\Sigma_{\sigma}(i\omega_n)$ allows us to compute the lattice Green's function \eqref{eq:lat_Green} or \eqref{eq:lat_Green_twosub}, depending on the type of ordering we have. Due to the self-consistency condition, the lattice Green's function substitutes the impurity Green's function in the Dyson equation \eqref{eq:Dyson}, thus updates the Weiss function and the corresponding Anderson parameters. The procedure is repeated until convergence.

In addition to a direct application of the DMFT self-consistency conditions related to Eqs.~\eqref{eq:lat_Green} or \eqref{eq:lat_Green_twosub}, in the low-temperature region we employ a real-space generalization of this method to the finite-size or inhomogeneous  lattice systems (RDMFT) \cite{Hel2008PRL,Sno2008NJP,Sotnikov2015PRA,unukovych20214}. While the general idea of the described method remains the same, the lattice Green's function is now obtained from the inversion of the real-space matrix,
\begin{eqnarray} \label{eq.Nsub}
    [{\bf G}^{-1}_{\sigma}(i\omega_n)]_{ jj'} 
    = [i \omega_{n} + \mu 
    - \Sigma_{\sigma}(i \omega_n)]\delta_{ jj'} - t_{ jj'},
\end{eqnarray}
where indices $j$ and $j'$ represent the lattice sites, $\delta_{jj'}$ is the Kronecker delta symbol, and $t_{jj'}$ is the matrix element corresponding to the tunneling amplitude between the sites $j$ and $j'$, which takes values $0$, $t$, or $t_z$, depending on the sites location. Note that we perform RDMFT calculations with the periodic boundary conditions. 

Within our study, depending on the value of anisotropy determined by the ratio $t_z/t$, two different real-space configurations of clusters \cite{Sotnikov2015PRA} in the lattice were employed. For the case of (almost) decoupled parallel planes ($t_z\ll t$), the impurity-solver calculations were performed on 32 sites, which are located in real space on two adjacent layers, each consisting of $4\times4$ plaquette (see also Ref.~\cite{unukovych20214}). The initial configuration on the second layer (plaquette) was obtained from the other by the translation by one site along crystallographic axis in the plane of the layer. In the opposite limit of anisotropy ($t_z=t$), we performed calculations on 8 sites, forming the vertices of the cube. For both cases, we perform translations along crystallographic axes to cover a system of at least $8\times8\times8$ sites, accounting for the translational symmetry of clusters, and perform DMFT calculations for different values of the anisotropy $t_z/t$. Although the system consisting of $8^3$ sites may seem relatively small, we verified that the converged results remain consistent with the ones for the systems of the size $12^3$. In the intermediate regimes of $t_z/t$, we performed calculations with both options for the cluster translations. Depending on the specific parameter regime (determined by the values of $U$, $T$, and $t_z$), we observe that only one or two configurations show stable convergence from the given sets of initial Anderson parameters. 

\section{Results}\label{sec.2}
\subsection{Magnetic phases and their dependence on anisotropy}
\begin{figure}[t]
    \centering
    \includegraphics[width=\columnwidth]{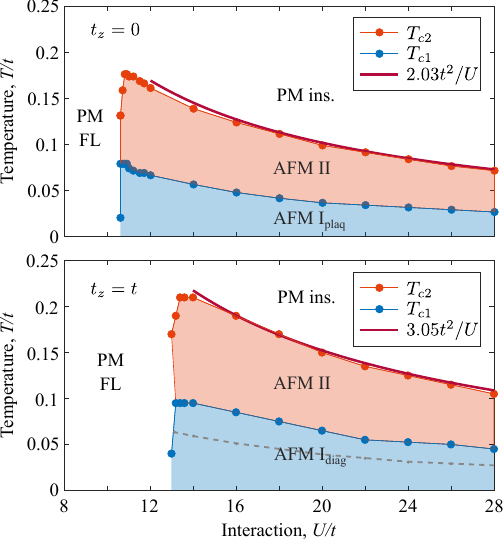}
    \caption{Phase diagrams of the SU(4)-symmetric Hubbard model at quarter filling ($n=1$) for a square lattice, i.e., decoupled layers (top) and isotropic cubic lattice (bottom). On both diagrams, bold solid lines correspond to the least-square fitting of the DMFT data points for the upper critical temperature $T_{c2}$ to $J\propto t^2/U$ (strong-coupling approximation). \change{In the lower panel, the dashed line inside AFM I$_{\rm diag}$ indicates the upper critical temperature $T_{c1}$ for the coexisting (metastable) phase AFM I$_{\rm plaq}$.}}
    \label{fig:phase_diag}
\end{figure}

First, we perform the numerical analysis to both calculate magnetic ordering for an isotropic cubic lattice ($t_z=t$) and perform additional computation for decoupled planes ($t_z=0$). The latter limit physically corresponds to the isotropic 2D square lattice, studied with DMFT for the SU(4)-symmetric Hubbard model in Ref.~\cite{unukovych20214}. 
Based on the converged results, we construct phase diagrams for the two limiting cases on the same scale (in units of the in-plane hopping amplitude~$t$), as shown in Fig.~\ref{fig:phase_diag}.  One can see that irrespectively of the interlayer coulping~$t_z/t$, the SU(4)-symmetric system at quarter filling demonstrates two antiferromagnetic (AFM) insulating regimes, as well as the paramagnetic (PM) phase with a smooth crossover from the Fermi-liquid to insulating behavior with the increase of $U$ at $T\gtrsim0.25t$. Albeit the hysteresis behavior can appear at the boundary of the AFM insulator and Fermi-liquid states, as well as between AFM phases with different residual symmetries (AFM I and AFM II), in this study we do not analyze this aspect in deep detail. Here, we depict the boundaries for the AFM regions corresponding to the maximal stability of the respective phase, which have the lower residual symmetry among the two phases.

Naturally, the two diagrams for two limiting cases look similar to each other, with the PM insulator phase for the isotropic cubic lattice (lower panel) appearing at larger amplitudes of the local interaction $U$ and higher temperatures $T$ than for the decoupled layers (upper panel). Note that the critical temperatures corresponding to the transition from AFM~II to PM in the strong-coupling regime are approximately related to each other by the factor $3/2$. This is not a coincidence and can be explained within the effective Heisenberg model. Naturally, in accordance with Eq.~\eqref{eq:Heis_Ham}, the energy of the local spin excitation must be proportional to both the effective magnetic coupling~$J$ and the lattice connectivity (number of the nearest-neighbor sites). The connectivities are 4 and 6 for the two lattice limits, respectively, and $T_{c2}\propto J\propto t^2/U$, thus one can perform approximate estimates by rescaling the energy-related quantities (in units of $t$) by the factor~$\sqrt{3/2}$.
Additionally, this fact also explains the reason why the critical couplings $U_c/t$ ($T\to0$) in the two limits are related to each other by the factor $\sqrt{3/2}$ ($U_c\approx10.8t$ and $U_c\approx13t$, respectively), but not by the factor $3/2$.

\begin{figure}[t]
    \centering
    \includegraphics[width=\columnwidth]{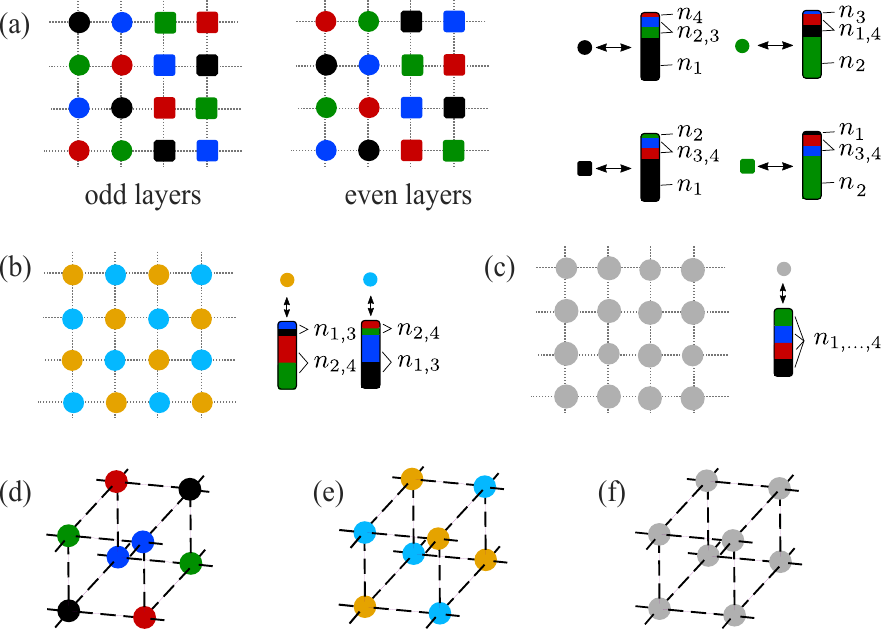}
    \caption{Spatial configurations of local occupancies on the layered square and cubic lattices: (a) AFM I plaquette configurations and (d) AFM I diagonal ordering. Two-sublattice ordering (AFM II) is shown in (b) and (e), while (c) and (f) describe unordered (PM) states in 2D and 3D configurations, respectively. 
    Color coding corresponds to the dominant local occupancy by a particular spin component, while shapes of symbols in (a) correspond to two permutations of minor occupancies.}
    \label{fig:lattice}
\end{figure}

The key difference in properties of the two geometries can be found in the regime $T\to0$, where the AFM~I ordering  emerges. The (quasi-)two-dimensional case supports the AFM plaquette ordering of spin components on the lattice, as shown in Fig.~\ref{fig:lattice}(a). Here, the different color filling of symbols denote the prevailing pseudospin component, meantime the shape describes which pseudospin components is the least probable. For instance, consider the sites denoted by green circles and green squares in Fig.~\ref{fig:lattice}(a). 
Let the green color correspond to the dominant component with $\sigma=2$.
On the sites denoted by green circles, the first and the fourth pseudospin components have the same occupancy, while on the sites denoted by green squares the third and the fourth pseudospin components are equiprobable. 
Although at $t_z=0$ the plaquette ordering on each layer can be chosen independently with respect to the square-lattice regular translations, axes reflections, and $C_4$ rotations, we specifically chose the same plaquette ordering on each odd layer and its translation by one site along one particular crystallographic axis on each even layer with respect to the odd layer.
As for the opposite case, the isotropic cubic lattice limit ($t_z=t$), the corresponding AFM~I (diagonal) ordering can be viewed as sets of parallel diagonals, where color represents the dominant pseudospin flavor on the site with three other flavors being equiprobable, as shown in Fig.~\ref{fig:lattice}(d).

\change{Although our DMFT analysis (restricted to diagonal observables in the pseudospin space) suggests different local occupancies $\langle\hat{n}_{i\sigma}\rangle$ for both types of the AFM I phases on the given lattice site, it can be rather challenging to distinguish AFM~I plaquette from AFM~I diagonal phase by analyzing only local observables under experimental conditions. 
The DMFT results rely on relative differences between the densities of the minority components, which typically do not exceed 0.05 in the AFM~I plaquette phase (and zero in the AFM~I diagonal phase).
By looking at the experimental progress with quantum gas microscope techniques \cite{Yamamoto16,Miranda17} it looks more straightforward to measure the nonlocal spin-spin correlation functions of the type $\langle \hat{S}_{ki}\hat{S}_{kj} \rangle$, which naturally show different behavior in these two phases.
}

With the temperature increase, the two-sublattice AFM~II ordering becomes thermodynamically stable, identically for both limits of the interlayer coupling. Here, the system shows regular bipartite modulations of pseudospin components divided into two pairs, as shown in Figs.~\ref{fig:lattice}(b) and \ref{fig:lattice}(e). With the further temperature increase, every site becomes equally occupied by every flavor on average, which we associate with the unordered or paramagnetic (PM) state, as shown in Figs.~\ref{fig:lattice}(c) and \ref{fig:lattice}(f).

\begin{figure}[t]
    \includegraphics[width=\columnwidth]{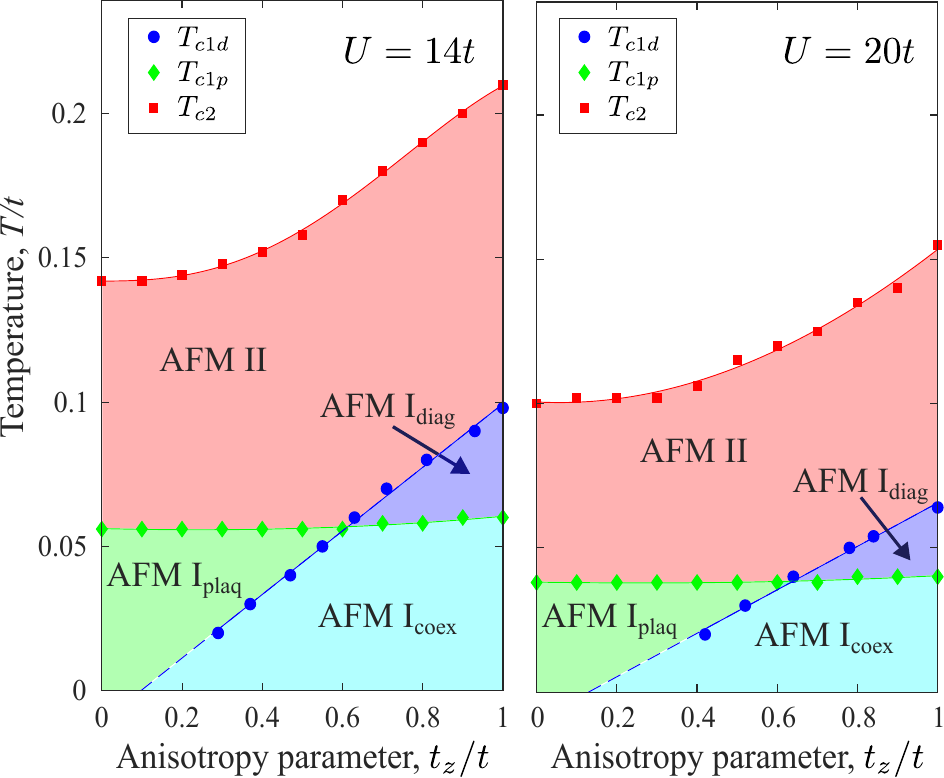}
    \caption{Magnetic phase diagrams at two different interaction strengths $U$, indicating stability regions and transition lines as functions of temperature and lattice anisotropy. Solid lines fit the data from DMFT calculations, while blue dashed lines are linear extrapolations in the limit $T\to0$.
    }
    \label{fig:temp_anis}
\end{figure}
Since the AFM~I plaquette (2D) and AFM~I diagonal (3D) phases differ in the way that one ordering cannot be continuously transformed into other one, the question of the transition (and the corresponding critical coupling~$t_z/t$) arises. We determine and plot the region of existence of magnetic phases depending on the temperature and anisotropy parameter $t_z/t$ for two different values of the interaction strength in Fig.~\ref{fig:temp_anis}. According to our DMFT analysis, at sufficiently low temperatures the plaquette ordering remains stable on the whole interval $t_z\in[0,t]$, with a slight increase of the critical temperature~$T_{c1}$ with the increase of $t_z$. At the same time, the critical temperature~$T_{c1}$ corresponding to the diagonal ordering shows almost linear dependence on $t_z$. The dashed part of the line shows the extrapolation of the dependence to $T\to0$, where the developed DMFT algorithm becomes less reliable. Intersection of both regions gives ranges of temperatures and spatial anisotropies of the optical lattice, where both types of orderings can coexist and emerge depending on the initial-state preparation and pre-thermalization in the actual system under study. From Fig.~\ref{fig:temp_anis} one can also note that with the increase of the interaction amplitude $U/t$ (by comparing left and right panels), the ordered phases shrink on the temperature scale (i.e, along vertical direction), which corresponds to the suppression of the AFM coupling $J\propto t^2/U$ and also agrees with Fig.~\ref{fig:phase_diag}.


\subsection{Real-space distributions in a harmonic optical trap}
The coherent laser beams forming optical lattice in cold-atom experiments usually have Gaussian spatial profiles of their intensity. The respective dipole potential in the center of the trap can be approximated well by a harmonic potential, by restricting to the first two terms in the corresponding Taylor series. 
Therefore, it becomes relevant to analyze the spatial distributions and stability of magnetic states at low temperatures in the harmonic optical trap with specific amplitude $V_j$ of the external potential in the Hamiltonian (\ref{eq:Hubbard_Ham}). For the convenience of theoretical analysis, the trapping amplitude $V_j$ can be expressed as follows:
\begin{align}
    V_j(r_j)=V\frac{r_j^2}{a^2},
\end{align}
where $a$ is the lattice spacing, while the curvature $V$ and the chemical potential $\mu$ in the Hamiltonian \eqref{eq:Hubbard_Ham} are chosen below to result in $n\approx2$ (Mott plateau) in the center and zero filling at the edges of the trap. 
\begin{figure}[t]
    \includegraphics[width=\columnwidth]{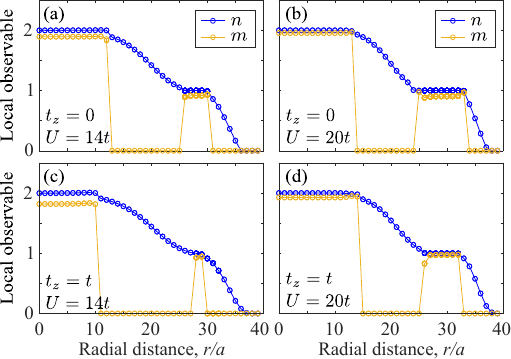}
    \caption{Distributions of local observables: atomic density~$n$ and order parameter~$m$ in the harmonic optical trap for different values of the interlayer hopping $t_z$ and interaction amplitude~$U$.  The parameters are $T=0.04t$ for $U=14t$ and $T=0.03t$ for $U=20t$.}
    \label{fig:loc_obs}
\end{figure}

To quantify the changes in magnetic ordering, we introduce a notation for the local filling of the lattice \change{site~$i$ by each spin component~$\sigma$}, where the index corresponds now to the dominance (and not fixed to a particular pseudospin component itself), i.e., we permute components in the way that it is always \change{$n_a\geq n_b\geq n_c\geq n_d\geq 0$ with $n_a=\max\langle\hat{n}_{i\sigma}\rangle$ and $n_d=\min\langle\hat{n}_{i\sigma}\rangle$, in particular}.
According to this notation, in Fig.~\ref{fig:loc_obs} we plot the radial distributions for both the total filling $n$ and the order parameter $m$,
\change{
\begin{align}
    &n=n_a+n_b+n_c+n_d,\nonumber\\
    &m=n_a+n_b-n_c-n_d.
\end{align}
}
The introduced positive-valued order parameter $m$ is nonzero in both types of the AFM phases; it has larger values in AFM II rather than in AFM I, and vanishes in the PM regime. We observe that at half-filling (Mott plateau with $n\approx2$) and in its small vicinity, the fermionic mixture can only have the two-sublattice solution among magnetically ordered states, which is also in agreement with Ref.~\cite{Golubeva2017PRB}. Away from the trap center, paramagnetic solutions (metallic states) take place at non-integer values of total filling $n$, while either plaquette or diagonal AFM phase emerges at another Mott plateau with $n\approx1$, depending on the interlayer coupling~$t_z/t$ in the lattice.

Let us also discuss two peculiarities in the depicted spatial distributions. First, in Fig.~\ref{fig:loc_obs}(b) in the dependence of the order parameter $m$ one can notice sharp features on both sides of the Mott plateau with  $n\approx1$, which correspond to the emergence of the stable AFM II solution (there, the AFM I solution becomes thermodynamically less stable due to not optimal external potential and nonzero $T$). Second, every density distribution contains step-like changes in $n(r)$, i.e., discontinuities in the derivative $\partial n/\partial r$, between the states with integer (Mott insulating) and non-integer values (metallic states) of $n$, which we attribute to the local-density approximation employed in the DMFT analysis. Note that within a more accurate (but more computationally demanding) full-size RDMFT approach for the trapped system, these features should be smoothed out, since it more accurately accounts for the proximity effects (see, e.g., Ref.~\cite{Sotnikov2012PRL} for comparison).

\section{Conclusion}
We studied the equilibrium magnetic properties of four-component SU(4)-symmetric interacting Fermi gas in a periodic lattice potential governed by the Fermi-Hubbard model at quarter filling for variable tunneling amplitude along one particular crystalographic axis. 
We employed the dynamical mean-field theory with the exact diagonalization impurity solver to determine emerging magnetic phases and corresponding critical regimes. In the two limiting cases of decoupled layers with square lattice geometry and isotropic cubic lattice we observe two types of AFM long-range ordered states and paramagnetic regimes. Independently on the value of the hopping anisotropy $t_z/t$, we established that AFM phases at quarter filling can emerge only in the strong-coupling regimes of the Hubbard model with $U\gtrsim11t$. 

The AFM phases with the lowest residual symmetry (AFM I plaquette and AFM I diagonal), which correspond to the limit $T\to0$, have different structures of magnetic ordering. These orderings cannot be linked to each other by a continuous transformation. This motivated us to study their stability under the change of the hopping anisotropy. 
Within the DMFT approach, we observed that the plaquette ordering remains stable in the low-temperature region covering the whole interval of $t_z\in[0,t]$, with a slight increase of the critical temperature with the increase of $t_z$. At the same time, the diagonal AFM ordering becomes significantly suppressed with a decrease of $t_z$, allowing both orderings coexist at the specific values of the system parameters. 
Under conditions of the experiments with ultracold atoms, this does not imply existence of both plaquette and diagonal orderings at the same time, but one of these can remain in the metastable state depending on the initial-state preparation and thermalization processes in the system. 

To address numerically the question of the exact ground state, more refined methods than DMFT are necessary, which is beyond the scope of the current study. In this respect, tensor-network approaches for the three-dimensional models \change{\cite{2021_Vlaar, Lukin2024spgap, Lukin2024ipeps3d}} seem to be the most promising candidates to determine the exact positions of the studied anisotropy-induced phase transition in the zero-temperature limit.

In addition, we theoretically analyzed spatial regions of stability of magnetic phases under conditions of the external harmonic confinement. This is a typical situation in the optical-lattice experiments for neutral atoms, where the effective harmonic confinement in the system center is produced by the orthogonal Gaussian beams forming optical trap. 
In the regions with two atoms per site (Mott plateau with $n\approx2$ or half filling), we observe stability of the two-sublattice AFM phase,  if the temperature (entropy) is sufficiently low. 
Further away, connected by the unordered metallic states with $2>n>1$, we observed another Mott plateau with $n\approx1$. There, the studied AFM I phase can emerge at sufficiently low temperature and $U>U_c$, with a possibility of narrow regions with  AFM II ordering appearing on the edges of the plateau.

\acknowledgements
The authors acknowledge support by the National Research Foundation of Ukraine under the call ``Excellent science in Ukraine'', project No. 235/0073 (2024--2026). V.U. also acknowledges financial support of STCU via the IEEE program ``Magnetism for Ukraine 2023'', Grant No. 9918.

\bibliography{Anisotropy}	
\end{document}